\documentclass[conference]{IEEEtran}
\IEEEoverridecommandlockouts
\usepackage{cite}
\usepackage{amsmath,amssymb,amsfonts}
\usepackage{algorithmic}
\usepackage{graphicx}
\usepackage{textcomp}
\usepackage{xcolor}
\usepackage{booktabs}
\usepackage{hyperref}
\usepackage{subfig}
\hypersetup{
colorlinks=true,
linkcolor=black,
citecolor=black
}
\usepackage[numbers,sort&compress]{natbib}
\def\BibTeX{{\rm B\kern-.05em{\sc i\kern-.025em b}\kern-.08em
    T\kern-.1667em\lower.7ex\hbox{E}\kern-.125emX}}
\begin{document}

\title{SourceP: Detecting Ponzi Schemes on Ethereum with Source Code}

\author{
\IEEEauthorblockN{Pengcheng Lu\textsuperscript{1},Liang Cai \textsuperscript{2}, Keting Yin\textsuperscript{1}}
\IEEEauthorblockA{
\textsuperscript{1}\textit{School of Software Technology, Zhejiang University, China} \\
\textsuperscript{2}\textit{College of Computer Science and Technology, Zhejiang University, China} \\
lupc@zju.edu.cn}
}

\maketitle

\begin{abstract}
As blockchain technology becomes more and more popular, a typical financial scam, the Ponzi scheme, has also emerged in the blockchain platform Ethereum. This Ponzi scheme deployed through smart contracts, also known as the smart Ponzi scheme, has caused a lot of economic losses and negative impacts. Existing methods for detecting smart Ponzi schemes on Ethereum mainly rely on bytecode features, opcode features, account features, and transaction behavior features of smart contracts, which are unable to truly characterize the behavioral features of Ponzi schemes, and thus generally perform poorly in terms of detection accuracy and false alarm rates. In this paper, we propose SourceP, a method to detect smart Ponzi schemes on the Ethereum platform using pre-trained models and data flow, which only requires using the source code of smart contracts as features. SourceP reduces the difficulty of data acquisition and feature extraction of existing detection methods. Specifically, we first convert the source code of a smart contract into a data flow graph and then introduce a pre-trained model based on learning code representations to build a classification model to identify Ponzi schemes in smart contracts. The experimental results show that SourceP achieves 87.2\% recall and 90.7\% F-score for detecting smart Ponzi schemes within Ethereum's smart contract dataset, outperforming state-of-the-art methods in terms of performance and sustainability. We also demonstrate through additional experiments that pre-trained models and data flow play an important contribution to SourceP, as well as proving that SourceP has a good generalization ability.
\end{abstract}


\section{Introduction}
Blockchain technology is rapidly evolving as an open ledger of recorded transactions maintained in a distributed network of mutually untrusted peers (i.e., peer-to-peer network), where each peer verifies transactions through a consensus protocol \cite{16, 3, 4, 5}. Blockchain has now developed applications in areas such as the Internet of Things \cite{17}, voting systems \cite{18}, authentication \cite{19}, ledgers \cite{4}, disaster relief \cite{20}, healthcare \cite{23}, and edge computing \cite{22}, which have attracted significant interest from industry and academia \cite{1}. Also, blockchain has become a common infrastructure for the emerging metaverse and Web3, and DeFi, DAO, and Non-Fungible Tokens developed from blockchain technology are becoming more and more popular \cite{24}.

Ethereum \cite{47} is currently the most popular public blockchain platform with smart contract functionality \cite{46}. Ethereum handles peer-to-peer contracts through the cryptocurrency Ether(ETH) and can store data, run smart contracts, and share information globally. First proposed by Nick Szabo in the mid-1990s \cite{14}, smart contracts contain contractual terms that will be automatically executed when predefined conditions are met, stored, replicated, and updated in a distributed blockchain. The combination of blockchain technology and smart contracts makes the dream of a ``peer-to-peer marketplace" come true, which means that there will be no third-party intervention in transactions between buyers and suppliers via smart contracts in a blockchain marketplace \cite{15}.

Smart contracts on the Ethereum platform are Turing-complete, and through Turing-complete smart contracts, users can not only trade in cryptocurrency but also perform arbitrary actions on the blockchain \cite{90}. Because of this, decentralized applications (DApps) \cite{88, 89} consisting of smart contracts are currently growing rapidly, such as CryptoKitties \cite{91} and IDEX \cite{92}. Yet at the same time, smart contracts have given a new opportunity for the spread of Ponzi schemes \cite{80}.

Born 150 years ago, Ponzi schemes are a classic financial fraud that lures new users in with the promise of high profits, with the core mechanism of compensating the investors and creators who joined before with the investments of new investors, creating no value themselves \cite{38}. Once no new investors join, the Ponzi scheme quickly collapses, new investors who are not compensated lose their money, and the scam makers and early participants receive illegal income. Smart contracts allow the creators of Ponzi schemes to remain anonymous, while their immutability makes them extremely difficult to change after being deployed to the blockchain, the feature that criminals exploit to deploy scams on smart contract platforms like Ethereum \cite{7}. Such Ponzi schemes deployed on smart contracts are called smart Ponzi schemes. Ponzitracker \cite{94} found close to \$3 billion amount of Ponzi schemes in 2022, related to blockchain-based cryptocurrencies. The recent smart Ponzi scheme event on the Forsage website involved more than \$300 million \cite{93}. The danger of financial fraud is so serious that reducing smart Ponzi schemes is imminent. Meanwhile, smart contracts are the building blocks of DApps, and if individual smart contracts in DApps are Ponzi schemes, then the DApp may also be dangerous. Therefore, detecting whether smart contracts deployed in Ethereum are smart Ponzi schemes and flagging them are important to prevent financial fraud and maintain the healthy development of DApps and blockchain platforms.



\begin{figure*}[ht]
    \centering

    \begin{minipage}[t]{0.49\textwidth}
        \centering
        \includegraphics[width=\textwidth]{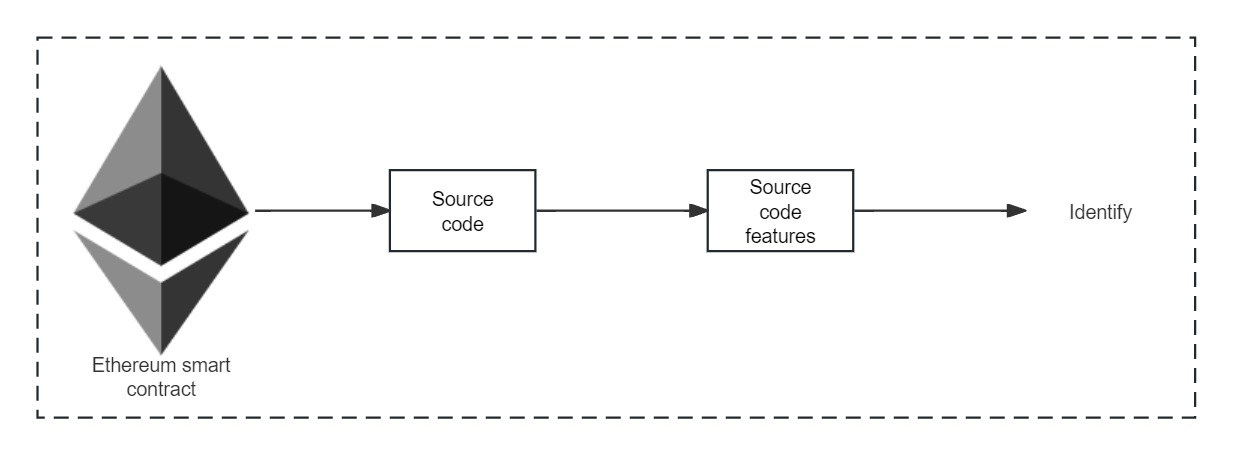}
        \caption{Flowchart of our method.}
        \label{fig:fig11}
    \end{minipage}
    \hfill
    \begin{minipage}[t]{0.49\textwidth}
        \centering
        \includegraphics[width=\textwidth]{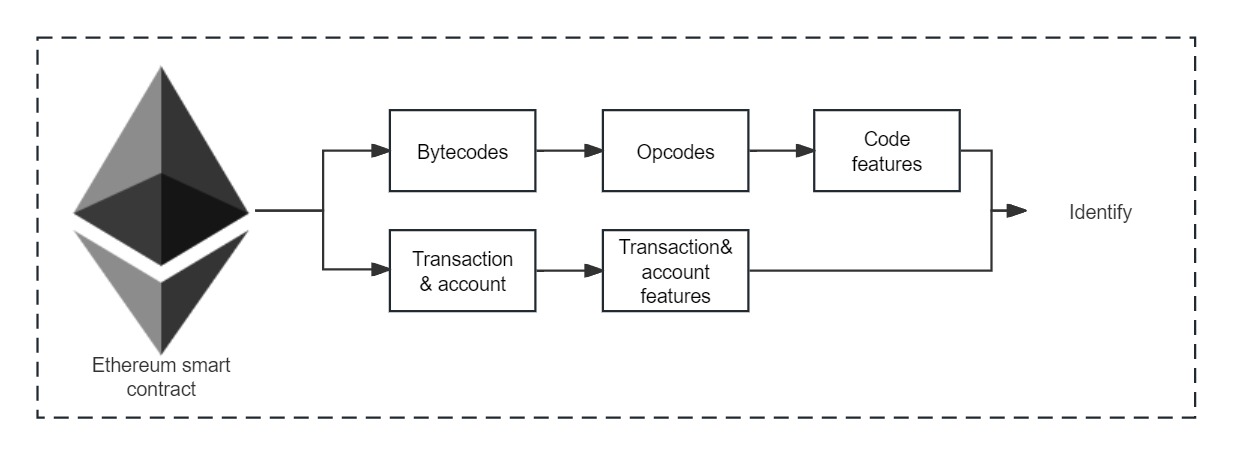}
        \caption{Flowchart of the traditional method.}
        \label{fig:fig12}
    \end{minipage}

\end{figure*}

To the best of our knowledge, the existing methods for detecting smart Ponzi schemes fall into three main categories; the first category of methods uses bytecode or opcodes from smart contracts as features to train classifiers or perform static analysis \cite{6,80}; the second category of methods uses the transaction behavior features of smart contracts \cite{97,102}; the third category of methods uses both opcodes features and account features to train the model \cite{5,42,98}. However, these three methods face some limitations. Firstly, bytecode and opcode features lack interpretability, and adding or removing some opcodes in smart contracts can be easily done to circumvent detection. Second, it is also difficult to accurately locate the fraudster on the anonymous Ethereum platform, so it is quite difficult to collect the fraudster's transaction behavior characteristics and account characteristics. Third, static analysis methods suffer from recognition lag and rely on expert rules. Finally, in the context of rapid updates to smart contracts, the sustainability and performance of currently available methods for detecting new smart Ponzi schemes have decreased significantly. So we consider using other features to detect Ponzi schemes, such as smart contract source code, which no one has considered before.

\textbf{Our Method.}
To address these challenges, we propose a smart Ponzi scheme detection approach based on pre-trained models and data flow called SourceP. Pre-trained models in current natural language processing (NLP) techniques have facilitated many tasks \cite{11,12,13}, thus, it is reasonable to explore the great potential of pre-trained models in smart Ponzi scheme detection. Since smart contracts on Ethereum are typically written in Solidity \cite{2,48}, SourceP will detect Ponzi schemes on smart contracts written in the Solidity language.

SourceP embodies a key innovation point, namely the use of only the source code of smart contracts as a feature. Figure \ref{fig:fig11} and Figure 
\ref{fig:fig12} show the significant differences between existing traditional methods and our method in detecting smart Ponzi schemes through flowcharts. As can be seen from the comparison of the two figures, our approach omits to obtain the bytecode, transaction, and account information of the smart contract, which reduces the amount of data required while avoiding the various problems associated with too many features. The specific implementation steps of SourceP can be seen in Figure \ref{fig:fig2} and Figure \ref{fig:fig3}. First, the smart contract source code is converted into a data flow, then the source code information is fed into the pre-trained model together with the data flow representing variable dependencies, and finally, the identification results of the smart contract Ponzi scheme are output and evaluated. Ultimately, we find that SourceP outperforms previous methods in smart Ponzi scheme detection.

The main \textbf{contributions} are summarized as follows:

\begin{itemize}
\item We propose a method, SourceP, to automatically detect Ponzi schemes in smart contracts. To the best of our knowledge, SourceP is the first method to detect smart Ponzi schemes using only the source code of smart contracts as features. It introduces a pre-trained model for learning code representations and uses the code semantic information provided by the data flow converted from the source code for detection.
\item We conducted extensive experiments to demonstrate that SourceP outperforms the state-of-the-art in detecting smart Ponzi schemes in terms of performance and sustainability under the same conditions, and has a good generalization ability.
\item We have exposed the dataset and source code\footnotemark of this study for other researchers to replicate our methods and evaluation to facilitate future research in this direction.
\end{itemize}

\footnotetext{https://github.com/Lpover/SourceP}

\section{Background}

\subsection{Ethereum and Smart contract}

Ethereum is the first open-source public blockchain platform that supports advanced and custom smart contracts with the help of a Turing-complete virtual machine called Ethereum Virtual Machine (EVM) \cite{85,86}. EVM is the environment in which smart contracts run, and each node in the Ethereum network runs an implementation of EVM and executes the same instructions. Smart contracts are written in languages such as Solidity and Serpent \cite{49}, and then the smart contract code is compiled into EVM bytecode and deployed on the blockchain for execution. Once the smart contract is deployed on the blockchain, the corresponding bytecode and creation transactions are permanently stored on the blockchain. Ethereum is now becoming the most popular platform for smart contract development and can be used to design various types of decentralized applications (DApps) that can be used for digital rights management, crowdfunding, and gambling \cite{46}. 

\subsection{Ponzi schemes on smart contract}
The ``Ponzi scheme" is a fraudulent investment scam that promises high rates of return with little risk to the investor. Ponzi schemes create returns for early investors by acquiring new investors. It is similar to a pyramid scheme in that it is based on using new investors' money to pay off early investors. Both Ponzi schemes and pyramid schemes eventually bottom out when the influx of new investors dries up and there is not enough money to turn around. Then such scams simply fall apart and all those who have not yet recouped their investments never get them back.
Smart contracts provide a perfect breeding ground for the modern fraudster. Whereas traditional financial fraudsters need to worry about the law, third-party institutions, and their public image, smart contracts don't have the same problems. 

Ponzi schemes on the blockchain have proliferated. Vasek et al. \cite{8} analyzed the supply and demand of bitcoin-based Ponzi schemes and identified 1780 different Ponzi schemes. Chainalysis \cite{9} investigated cryptocurrency crimes from 2017 to 2019 and found that 92\% of cryptocurrency scams were Ponzi schemes. Ponzi schemes on Ethereum are packaged as investment projects or gambling games that promise huge returns to those who invest. Some large-scale Ponzi schemes even build sophisticated websites and run aggressive marketing campaigns to attract investors \cite{93,103}. It is difficult for investors with little knowledge of blockchain to distinguish the true nature of smart contracts that are disguised as high-yield investment schemes \cite{6}.

Bartoletti et al. \cite{7} found that from August 2015 to May 2017, 191 smart Ponzi schemes active on Ethereum had collected nearly \$500,000 from more than 2,000 different users. They specifically analyzed 184 real smart Ponzi schemes on Ethereum and classified smart Ponzi schemes into four types, chained-shaped,  tree-shaped, waterfall as well and handover, most of which are chained-shaped with 82\% of the total, while the other types have only 2\% of the total and the rest are The remaining are unclassified other types.

\subsection{Smart Ponzi Scheme Case}
\begin{figure}[htbp]
\centerline{\includegraphics[scale=0.23]{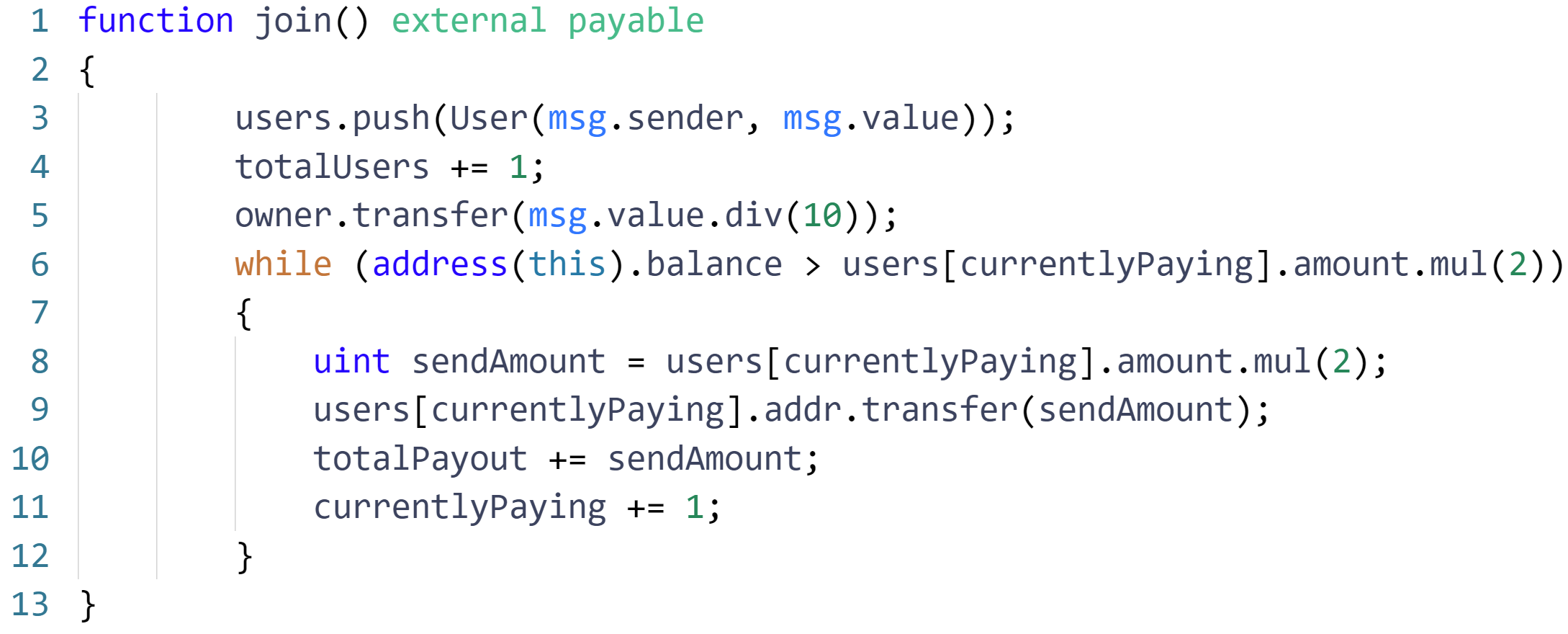}}
\caption{Source code fragment of a smart Ponzi scheme.}
\label{fig:fig1}
\end{figure}

Figure \ref{fig:fig1} shows an example of a smart Ponzi scheme, a Chain-shaped scheme of smart Ponzi source code fragment written by Roan \cite{25}. Each time a new investor participates in this smart Ponzi scheme, the $join()$ function is called and then adds the new user to the list of investors with their investment amount, and then transfers 10\% of the investment amount to the owner of the smart contract. Then, in a first come, first served order, if the total amount of the pool is already more than twice the amount invested by the first investor who has not been repaid, then he is compensated with twice the amount invested in Ether.

Roan said of the Ponzi scheme, ``That’s a complete lie, but enough to sucker someone in." \cite{26}

\subsection{Data Flow}
The data flow is a graph \cite{33}, which is also called a data flow graph (DFG), and it is a graph that represents the dependencies between variables in a code, where the nodes represent the variables and the edges represent the sources of the values of each variable. Data flow graphs are very useful in code analysis tasks \cite{27,28,29}. Unlike abstract syntax tree (AST), data flow has the same structure under different abstract syntaxes of the same source code, and such a structure provides critical code semantic information for code understanding. Moreover, dataflow has a simpler structure compared to AST, which is more efficient when used in models. So SourceP will use the dataflow graph as the input to the model. How to extract the dataflow from the source code will be described in Section \ref{DFGG}.

\subsection{Pre-trained model}
Pre-trained models can benefit a variety of downstream tasks by storing knowledge into huge parameters and fine-tuning them on specific tasks, the rich knowledge implicit in huge parameters has been extensively demonstrated through experimental validation and empirical analysis \cite{69}. Pre-trained models such as BERT \cite{34}, ELMo \cite{82}, GPT \cite{70}, and XLNet \cite{83} have been successful on various tasks. Serval pre-trained models for learning code representations are being applied, such as CodeBERT \cite{37}, CuBERT \cite{75}, GPT-C \cite{74}, and Code-GPT \cite{76}. SourceP uses GraphCodeBERT \cite{33}, a pre-trained model for learning code representations trained on the CodeSearchNet \cite{84} dataset, as the main part of the model. The specific method of the pre-trained model used by SourceP will be described in Section \ref{OM}.


\begin{figure*}[htbp]
  \centering
  \subfloat[Partial source code in a smart Ponzi scheme]
  {\includegraphics[width=0.62\textwidth]{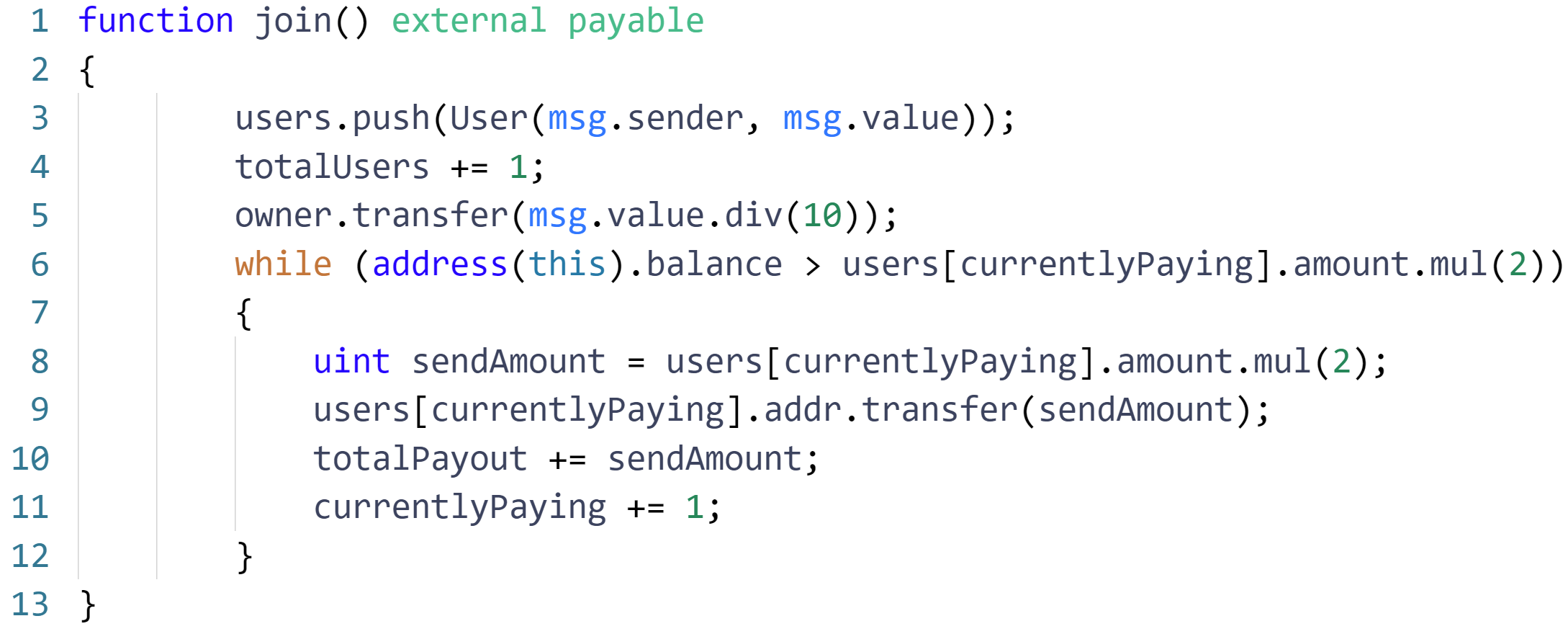}\label{fig:fig2a}}
  \subfloat[Parse into AST]
  {\includegraphics[width=0.37\textwidth]{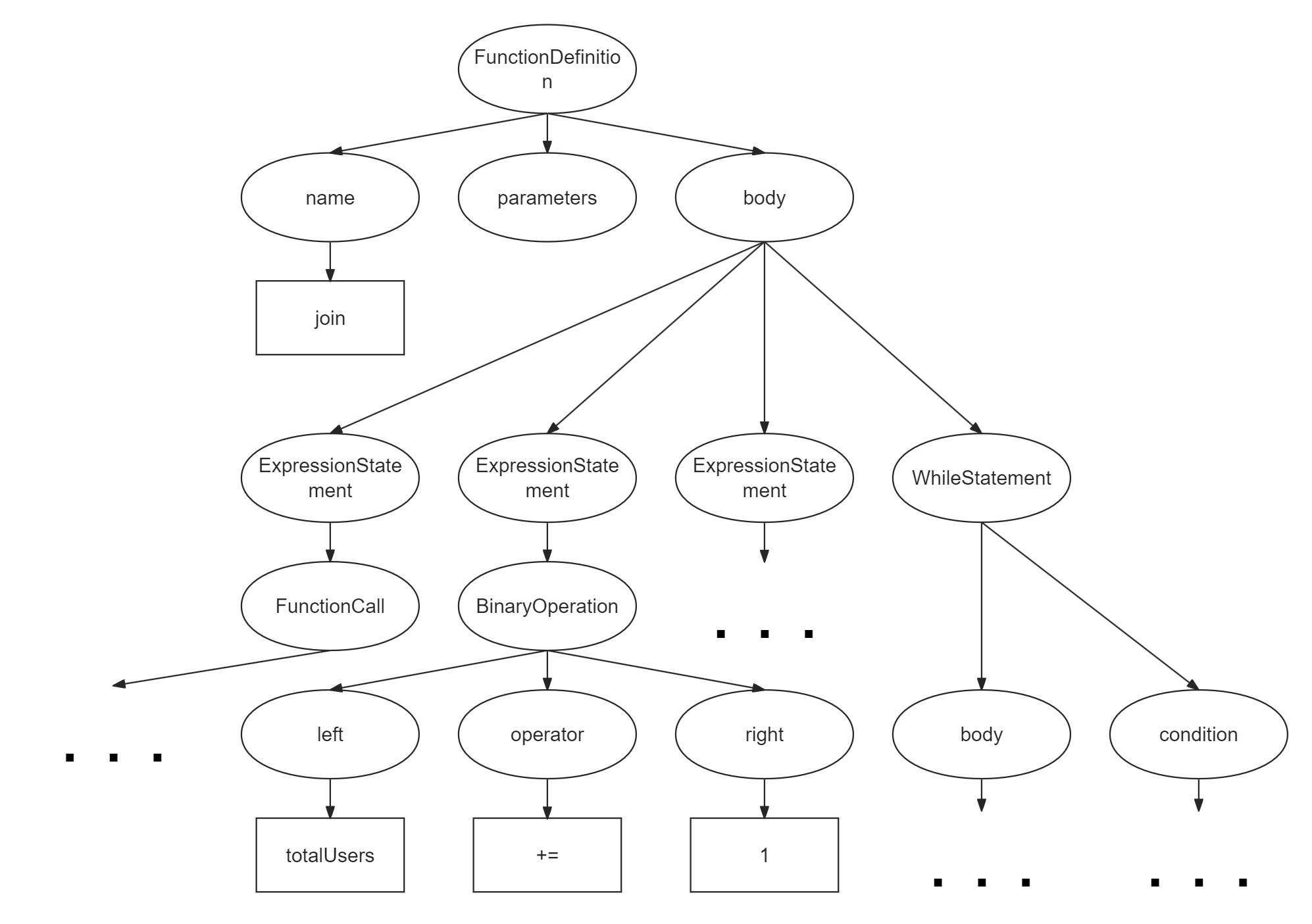}\label{fig:fig2b}}
  \\
  \subfloat[Identify variable sequence in AST]
  {\includegraphics[width=0.62\textwidth]{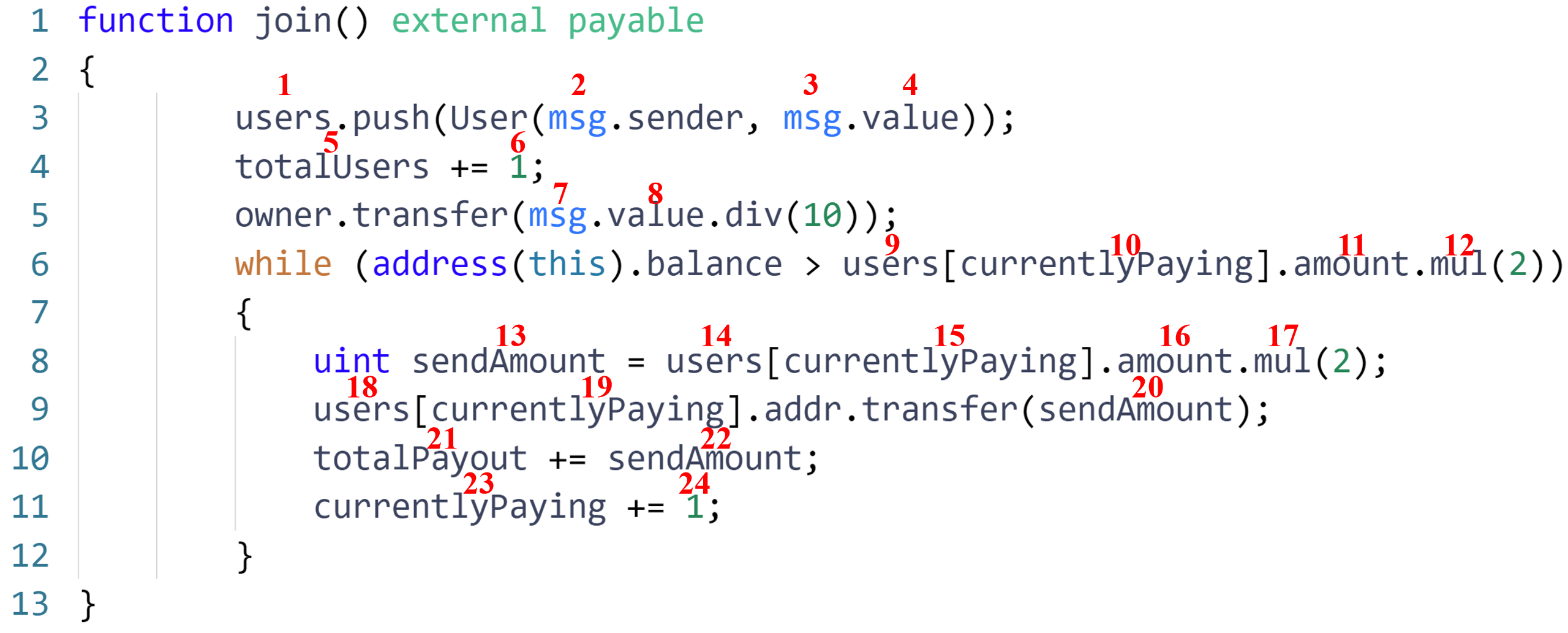}\label{fig:fig2c}}
  \subfloat[Data flow graph]
  {\includegraphics[width=0.37\textwidth]{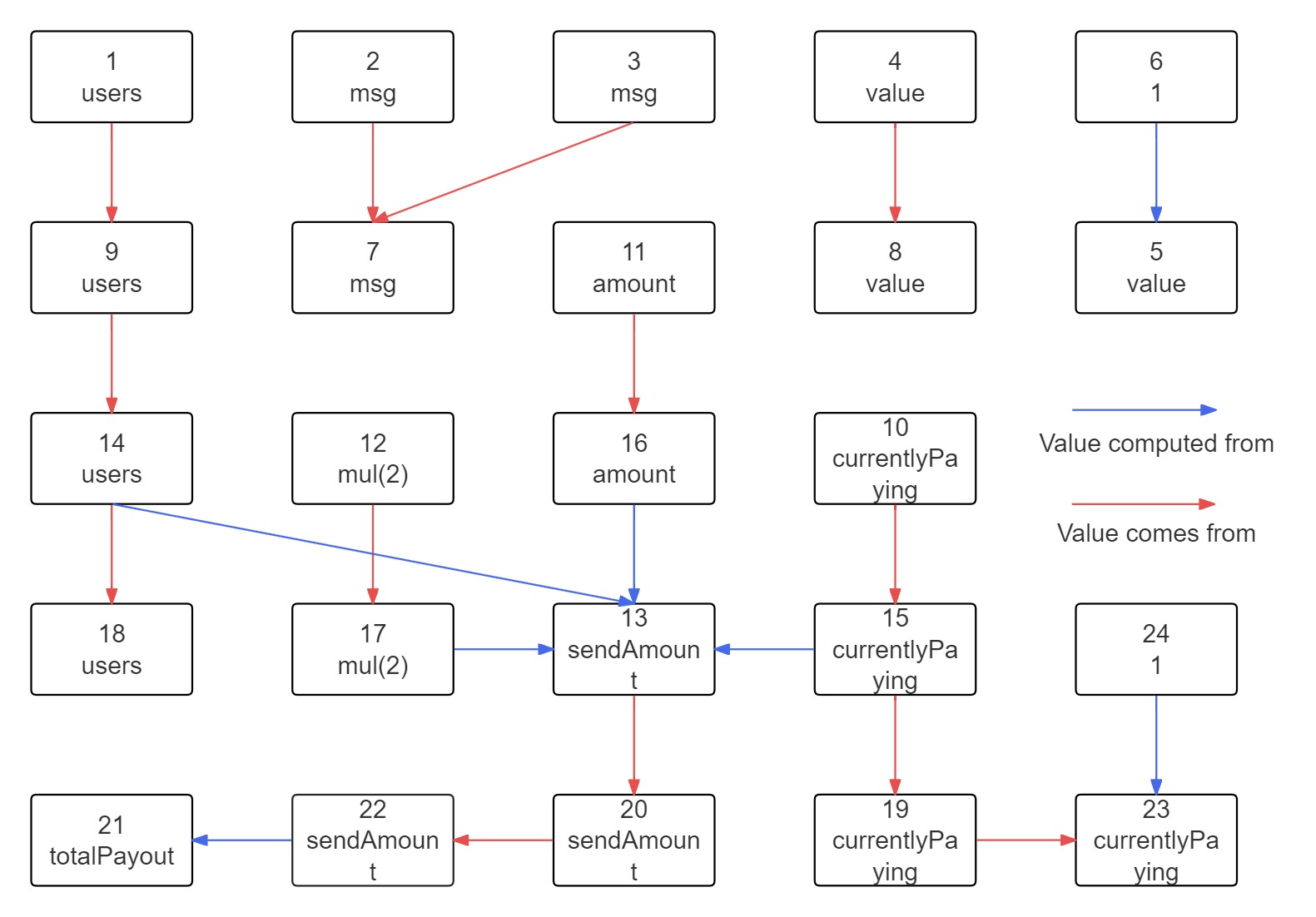}\label{fig:fig2d}}
  \\
  \caption{The phase of input normalization, i.e., the process of converting source code into data flow. (a) source code fragment of smart Ponzi scheme; (b) conversion of source code into abstract syntax tree (AST); (c) identification of variable sequences in abstract syntax tree (AST); (d) generation of data flow graph by variable sequences.}\label{fig:fig2}
\end{figure*}

\section{OUR METHOD}\label{OM}
\textbf{Method overview.}
Our method is divided into two main phases: 1) the input normalization phase, which converts the source code of smart contracts into AST and DFG; 2) the smart Ponzi scheme detection phase, which feeds the source code and DFG into a pre-trained model and outputs the final detection results. In the following, the first two subsections describe the details of each phase in detail. The next subsection deals with the functions used to incorporate the DFG into the Transformer. The last subsection is a detailed description of the three pre-training tasks of the pre-trained model.

\subsection{Data Flow Graph Generation}\label{DFGG}
\textbf{Source code to AST.}
First, we have the source code $SC=\left\{sc_{1}, sc_{2}, \ldots, sc_{n}\right\}$, then we convert the source code into abstract syntax trees (ASTs). Here we need to use a tool called tree-sitter \cite{30}, which can build a source code file into an AST. However, the tree-sitter does not provide official Solidity language support, so we need to use tree-sitter-solidity \cite{31} to convert the Solidity language source code to AST. This tool contains a grammar for tree-sitter which major inspiration and some structures have been taken from tree-sitter-javascript \cite{32}. AST includes the syntax information of the code, and the leaf nodes in the tree are used to identify the sequence of variables that are denoted as $Var=\{v_1,v_2,...,v_n\}$.

\textbf{AST to DFG.}
The data flow is a graph, so we can think of each variable in the AST as a node. The edge connecting two nodes is denoted as $\varepsilon=\left\langle v_{i}, v_{j}\right\rangle$, which means that the value of the $j$-th variable comes from the $i$-th variable. The set of all directed edges in the graph is denoted as $Edge=\left\{\varepsilon_{1}, \varepsilon_{2}, \ldots, \varepsilon_{n}\right\}$. So the final data flow graph is represented as $\mathcal{G}(SC)=(Var, Edge)$, this is the DFG we have constructed to represent the dependencies between variables of the source code. Figure \ref{fig:fig2} shows the process of converting the smart Ponzi scheme source code to AST and DFG.

\subsection{Model Structure}
\begin{figure}[htbp]
\centerline{\includegraphics[scale=0.23]{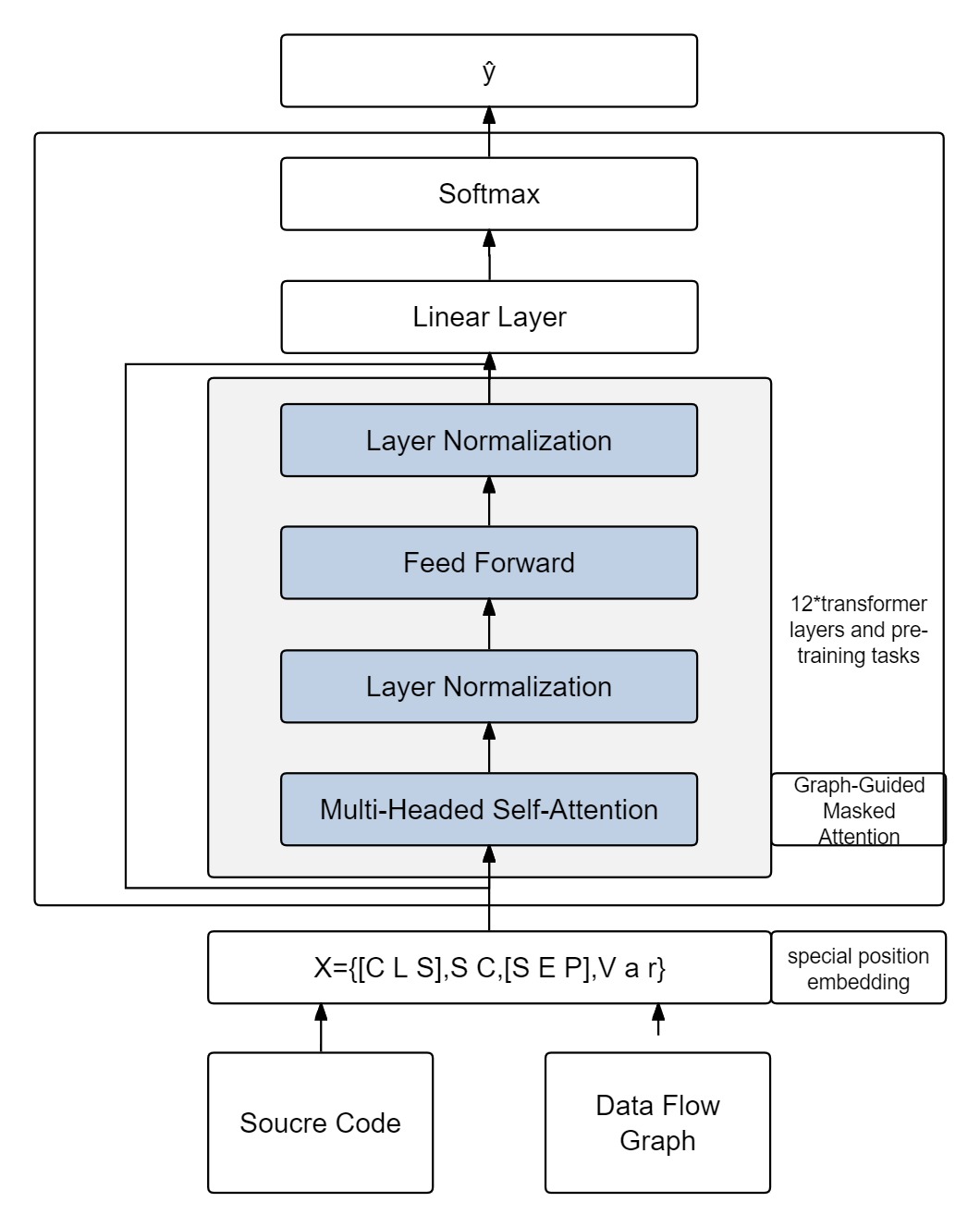}}
\caption{SourceP's model structure.}
\label{fig:fig3}
\end{figure}

In this section, we will describe the model structure of SourceP in detail. Our method primarily follows GraphCodeBert \cite{33}, so the model architecture follows BERT \cite{34} and the multi-layer bidirectional Transformer \cite{35} is the backbone of the model. Figure \ref{fig:fig3} shows the structure of the whole model.

The input to the model is the data flow graph converted from the source code, since the detection task is different, the paired comments of the input GraphCodeBert do not need to be considered. We modified the input by following Peculiar's \cite{10} method. The difference is that Peculiar uses a crucial data flow graph (CDFG) that retains only the data flow of key nodes as input to the model, but we still choose to use the DFG as input to the model. This is because Peculiar is designed to detect reentrant vulnerabilities in smart contracts and only cares about the dependencies of functions related to reentrancy vulnerabilities, while CDFG can remove redundant data flow information. But the case of smart Ponzi schemes is more complex and it may lurk in any function and variable, so it is necessary to use DFG as the input to the model.

So the input to the model is a sequence $X=\{[CLS], SC,[SEP], Var\}$. Where $[CLS]$ is a special classification token and $[SEP]$ is a special token used to split the two different data types. The remaining two segments in $X$, $SC=\left\{sc_{1}, sc_{2},..., sc_{n}\right\}$ is the collection of source code and $Var=\{v_1,v_2,...,v_n\}$ is the set of variables of data flow graph $\mathcal{G}(SC)=(Var,Edge)$.

Then the sequence $X$ will be converted to the input vector $W^{0}$. For each token in $X$, the corresponding token and position embedding are summed to construct its input vector. We use a special position embedding for all variables in $X$ as a way to indicate that these variables are nodes in the data flow. The model then uses the input vector $W^{0}$ to contextualize the representation through $N$ transformer layers $W^n=transformer_n(W^{n-1}),n\in[1,N]$. In our model, the value of $N$ is 12. The construction of each transformer layer is the same, and $U^n$ is obtained by first applying a multi-headed self-attention operation\cite{35} and then applying a Layer Normalization operation. The feed-forward layer and a Layer Normalization operation are then used on $U^n$. This way we get the output $W^n$ of the $n$-th layer from the input $W^{n-1}$. Here is the calculation for each transformer layer.

\begin{equation}
U^n=LayN(MulA(W^{n-1})+W^{n-1})
\end{equation}

\begin{equation}
W^n=LayN(FeeF(U^n)+U^n)
\end{equation}

Where $LayN$ means a Layer Normalization operation, $MulA$ means a multi-headed self-attention mechanism and $FeeF$ is a two-layer feed-forward network. In the $n$-th transformer layer, the multi-headed self-attention operation computes the $\hat{U}^n$.

\begin{equation}
Q_i=W^{n-1}P_i^Q,\:K_i=W^{n-1}P_i^K,\:V_i=W^{n-1}P_i^V
\end{equation}

\begin{equation}
head_i=Softmax(\dfrac{Q_i{K_i}^T}{\sqrt{d_k}}+M)V_i\label{eq4}
\end{equation}

\begin{equation}
\hat{U}^n=[head_1;...;head_m]{P_n}^O
\end{equation}

Where the output $W^{n-1}\in\mathbb{R}^{|X|\times d_h}$ of the previous layer is linearly projected onto a triplet of queries, keys, and values using the parameters $\begin{aligned}P_i^Q\text{,}P_i^K\text{,}P_i^V\in\mathbb{R}^{d_h\times d_k}\end{aligned}$ and $d_k$ is the dimension of a head. $M\in\mathbb{R}^{|I|\times|I|}$ is a mask matrix, $M_{\textit{ij}}$ is 0 if the $i$-th token is allowed to participate in the $j$-th token, otherwise it is $\text{-}\infty$. And $P_n^O\in\mathbb{R}^{d_h\times d_h}$ is the model parameters.

The model finally outputs the predicted label $\hat{y}$ through a linear classifier and the $Softmax$ function.

\begin{equation}
\hat{y}=\textit{Softmax}(\hat{U}^n)
\end{equation}

\subsection{Graph-guided masked attention}

In order to incorporate the structure of the data flow graph into the Transformer, GraphCodeBERT \cite{33} proposed this function. The masked attention function avoids the key $k_{i}$ that the query $q_{j}$ focuses on by changing the attention score ${q}_j^T k_i$ to $\text{-}\infty$ so that the attention weight becomes 0 after using the softmax function. To represent dependencies between variables, a node-query $q_{v_{i}}$ is allowed to attend to on a node-key $k_{v_{j}}$ if there is a direct edge from node $v_j$ to node $v_i$ where $\left\langle v_{j},v_{i}\right\rangle\in Edge$ or if $i=j$. otherwise, attention is masked by adding $\text{-}\infty$ to the attention score. To represent the relationship between source code tokens and data flow nodes, we first define a set $Edge^{'}$ where $\langle v_i,sc_j\rangle/\langle sc_j,v_i\rangle\in Edge^{'}$ if the variable $v_i$ is identified from the source code token $sc_j$. Then, we allow node $q_{v_i}$ and code $k_{sc_{j}}$ to attend each other when and only when $\langle v_i,sc_j\rangle/\langle sc_j,v_i\rangle\in Edge^{'}$. We use the following graph-guided masked attention matrix as the mask matrix $M$ in \eqref{eq4}.

\begin{equation}
M_{ij}=\begin{cases}0&\text{if }~q_i\in\{[CLS],[SEP]\}\\ &\text{or}~q_i,k_j\in P\cup SC\\ &\text{or}~\langle q_i,k_j\rangle\in Edge\cup Edge'\\ -\infty&\text{otherwise}\end{cases}
\end{equation}

\subsection{Pre-Training Tasks}
\textbf{Masked Language Modeling.}
The task follows Devlin et al.\cite{34} to apply a Masked Language Modeling (MLM) pre-training task. In particular, we randomly sampled 15\% of the tokens from the source code and paired annotations. We replace them with a [MASK] token 80\% of the time, replace them with random tokens 10\% of the time, and leave them unchanged 10\% of the time. The goal of MLM is to predict the original tokens of these sampled tokens, which is effective in previous work \cite{34,36,37}. In particular, if the source code context is not sufficient to infer the masked code tokens, the model can make use of the annotation context, encouraging the model to unify the natural language and programming language representations.

\textbf{Data Flow Edges Prediction.}
The purpose of this task is to learn representations from the data flow. The motivation is to encourage models to learn structure-aware representations that encode ``where the value comes from" relationships in order to better understand the code. In particular, we randomly extract 20\% of the nodes $Var_s$ in the data flow, mask the direct edges connecting these extracted nodes by adding $\text{-}\infty$ to the masking matrix, and then predict the $Edge_{\textit{mask}}$ of these masked edges. Formally, the pre-training objective of the task is calculated as \eqref{eq8}, where $Edge_{\textit{sc}}=Var_s\times Var\cup Var\times Var_s$ is the set of candidates for edge prediction, $\delta(e_{ij}\in Edge)$ is 1 if $\left\langle v_{i},v_{j}\right\rangle\text{}\in Edge$ otherwise 0. The probability $p_{e_{ij}}$ of the existence of an edge from the $i$-th node to the $j$-th node is calculated by dot product according to the $Sigmoid$ function, using the representation of two nodes in the model. To balance the positive-negative ratio of the examples, we sampled the same number of negative and positive samples of $Edge_{\textit{mask}}$.

\begin{equation}\label{eq8}
\begin{split}
loss_{EdgePred}=-\sum_{e_{ij}\in Edge_{\textit{sc}}}[\delta(e_{ij}\in Edge_{mask})log p_{e_{ij}}\\ +(1-\delta(e_{ij}\in Edge_{mask}))log(1-p_{e_{ij}})]
\end{split}
\end{equation}

\textbf{Node Alignment.}
The purpose of this task is to align the representation between source code and data flow, which is similar to data flow edge prediction. Instead of predicting the edges between nodes, we predict the edges between code tokens and nodes. The motivation is to encourage the model to align variables and source code to the data flow.

\section{Experiments}
To evaluate our method, we designed experiments to answer the following research questions (RQs):
\begin{itemize}
\item RQ1: How well does SourceP perform in detecting smart Ponzi schemes by relying only on source code features, and how do its precision, recall, and F-score compare to current state-of-the-art methods?
\item RQ2: How sustainable is SourceP in detecting smart Ponzi schemes?
\item RQ3: How do SourceP's pre-training tasks and data flow affect detection results?
\item RQ4: How well does SourceP generalization ability in detecting smart Ponzi schemes?
\end{itemize}

\subsection{Experiments Setting}
\textbf{Datasets.}
We used the Ponzi Contract Datasets provided by the XBlock platform \cite{5}, a dataset that crawls 6,498 smart contracts on Etherscan \cite{87}, and manually read the source code pairs to classify them with reference to the methods used in some previous studies \cite{7,42}, of which 318 smart contracts were manually marked as Ponzi smart contracts and the rest were manually marked as non-Ponzi smart contracts. We retained the source code in the dataset, along with the corresponding serial number of idx and the label. If the value of the label is 1 it means it is a smart Ponzi scheme, if it is 0 it is not.

\textbf{Evaluation metrics.}
For the experiments in the paper, we will use common precision, recall, and F-score to evaluate the performance of the model.

\textbf{Parameter settings.}
In the fine-turning step, we set the code length to 256, data flow length to 64, train batch size to 1, eval batch size to 32, learning rate to 2e-5, and used the Adam optimizer to update the model parameters. Depending on the experiment, epochs were set to 3, 5, or 10. Because the dataset has fewer positive category samples in some tasks, we set the thresholds from 0.5 to 0.15 and 0.003, which allows the model to be more inclined to predict positive samples that would otherwise be judged as negative categories.

\textbf{Implementation details.}
All the experiments were run on a computer with two Intel(R) Xeon(R) Silver 4314 CPUs at 2.4GHz, one GPU at NVIDIA A40 or two GPUs at NVIDIA A30, and 256GB of Memory.

\subsection{Results Summary}
\textbf{RQ1: Performance comparison with state-of-the-art methods.}
We compare the detection performance of SoucreP with the same division on the same dataset as the existing state-of-the-art method according to the comparison method of Zheng et al.\cite{5}. Specifically, all contracts are ranked according to the order of the block height at the time of smart contract creation, and the training set consists of smart Ponzi schemes from the 1st to the 250th and non-Ponzi smart contracts in between. The test set consists of the 251st to the last 341st smart Ponzi scheme and the remaining non-Ponzi scheme smart contracts. Thus, the training set has a total of 5990 smart contracts, while the test set has 508 smart contracts. Such a division, compared to random division, provides a better representation of the model's ability to detect emerging new smart Ponzi schemes when it only has data on earlier smart Ponzi schemes. The compared models include Ridge-NC \cite{5}: a ridge classifier model trained with an N-gram count feature; SVM-NC \cite{5}: an SVM model trained with an N-gram count feature; XGBoost-TF-IDF \cite{5}: an XGBoost model trained model with TF-IDF feature; MulCas \cite{5}: a multi-view cascade combinatorial model; SadPonzi \cite{80}: a semantic-aware system for detecting smart Ponzi schemes. The first three approaches use features extracted from the opcode of a smart contract, Mulcas incorporates Developer Feature on top of that, while SadPonzi detects Ponzi schemes based on the bytecode of a smart contract. The comparison results of different methods are listed in Table \ref{tbl:table1}. As can be seen from the results, SourceP shows the best performance in all three metrics. In particular, SourceP gains a 21.3\% recall improvement and a 12.9\% F-score improvement over the state-of-the-art method while ensuring that precision is also improved. Since the ratio of positive and negative samples is about 1:20, it is reasonable that the model prefers to classify the minority samples as the majority, resulting in a relatively higher precision score than the recall score.

\begin{table}
\centering
\caption{Comparison of SourceP on recall, precision, and f-score on a fixed training and test set with state-of-the-art methods}
\begin{tabular}{llllll} 
\toprule
\textbf{Method} & \textbf{Recall} &  \textbf{Precision} &  \textbf{F-score} \\
\midrule
XGBoost-TF-IDF & 0.234 & 0.882 & 0.370 \\
SadPonzi &  0.52 & 0.59  & 0.55 \\
SVM-NC & 0.375 & 0.923  & 0.533 \\
Ridge-NC &   0.453 &  0.829  & 0.586 \\
MulCas &  0.674 & 0.951  &  0.789 \\
\midrule
\textbf{SourceP} &  \textbf{0.887} & \textbf{0.956}  & \textbf{0.918} \\
\bottomrule
\end{tabular}
\label{tbl:table1}
\end{table}

\begin{table}
\centering
\caption{Comparison of SourceP on the three metrics on the dataset divided by block creation height with state-of-the-art methods}
\begin{tabular}{llllll} 
\toprule
\textbf{Method} & \textbf{Metric} &  \textbf{P2} &  \textbf{P3} & \textbf{P4} & \textbf{P5}\\
\midrule
    &  Precision &  0.33  & 0.42 & 0.18 &  0.24 \\
SadPonzi & Recall & \textbf{1.0}  &  0.71 & 0.25 & 0.18\\
 & F-score & 0.5 & 0.53 & 0.21 & 0.20 \\
\midrule
    &  Precision  &  0.88 &  0.96  & 0.81 &  0.95 \\
MulCas & Recall &  0.38 & 0.32 &  \textbf{0.94} &  0.67\\
    & F-score &  0.53 &  0.48 & 0.87 & 0.79\\
\midrule
    &  Precision  &  \textbf{0.99} &  \textbf{0.97}  & \textbf{0.88} &  \textbf{0.96} \\
\textbf{SourceP} & Recall &  0.55 & \textbf{0.89} &  0.90 &  \textbf{0.89}\\
    & F-score &  \textbf{0.59} &  \textbf{0.92} & \textbf{0.88} & \textbf{0.92}\\
\bottomrule
\end{tabular}
\label{tbl:table2}
\end{table}

\textbf{RQ2: Sustainability of the model compared to other state-of-the-art methods.}
Although SourceP has achieved excellent performance on the latest smart Ponzi schemes detection, there is a new problem known as model aging that has attracted widespread attention \cite{95,96}. In particular, there is a big difference between the early smart Ponzi schemes and the latest Ponzi schemes in smart contracts \cite{5,7}. To verify the sustainability of SourceP, in this experiment, we divide the dataset into six parts (P0 to P5) according to the height of the created blocks of Ponzi schemes, following the method of Zheng et al. \cite{5} divide the dataset by every 50 smart Ponzi schemes, e.g. block height in the first 50 smart Ponzi schemes and the non-Ponzi contracts among them are P0, while the smart Ponzi schemes with heights from 51 to 100 and the non-Ponzi contracts among them are divided into P1, and so on for the remaining P2, P3, P4, and P5. where P5 is the 251st to the last (314th) smart Ponzi scheme and the non-Ponzi contracts among them. P5 is actually the experiment of RQ1. The detection task is to predict the next dataset using the previous dataset, e.g., P0 and P1 to predict P2, and P0, P1, and P2 to predict P3. Because smart contracts are tamper-evident, a lower block height at creation means an earlier creation time, which is equivalent to our use of an earlier time smart Ponzi scheme to predict future smart Ponzi schemes as a way to verify the sustainability of SourceP. Our comparison models include SadPonzi \cite{80} and MulCas \cite{5}, and the results are shown in Table \ref{tbl:table2}. It can be seen that SourceP obtains the highest precision and F-score in each part of the experiment, achieving the highest recall in P3 and P5, and the F-score even improves by 39\% when detecting the P3 part. Since the new smart Ponzi scheme is deployed based on the ERC-20 token trading contract, the reward of the Ponzi scheme is reflected in the increased value of the Ponzi tokens, and SadPonzi and MulCas experience a degradation in their performance in detecting this new smart Ponzi scheme. This experiment demonstrates that SourceP achieves the best sustainability in detecting smart Ponzi schemes.

\textbf{RQ3: Ablation experiments.} We conducted ablation experiments to explore the contribution played by the pre-training task and the data flow in the smart Ponzi scheme detection by removing two pre-training tasks and the data flow respectively and then performing the task of RQ1 to observe the final results. The results are shown in Table \ref{tbl:table3}, where $-w/o$ is an abbreviation for without. From the results, we can see that removing the two pre-training tasks or not using the data flow makes the performance of the model degrade to different degrees, which indicates that they play a role in improving the model.

\begin{table}
\centering
\caption{Comparison of ablation experiments on the three metrics with pre-training tasks or data flow removed}
\begin{tabular}{llllll} 
\toprule
\textbf{Method} & \textbf{Recall} &  \textbf{Precision} &  \textbf{F-score} \\
\midrule
SourceP &  \textbf{0.887} & \textbf{0.956}  & \textbf{0.918} \\
-w/o EdgePred &   0.867 &  0.919  & 0.891 \\
-w/o NodeAlign & 0.821 & 0.914  & 0.860 \\
-w/o Data Flow & 0.806 & 0.909 & 0.847 \\
\bottomrule
\end{tabular}
\label{tbl:table3}
\end{table}

\textbf{RQ4: Generalization ability of SourceP.}
To verify the generalization ability of the model, we will divide the data set randomly to experiment. Also, to prevent model overfitting, we divide the validation set separately from the training set. Therefore, the ratio of training set: validation set: test set in this experiment is 7:1:2. We show the average results of 20 experiments in Table \ref{tbl:table4}. As we can see from the table, SourceP has good performance and generalization ability in detecting smart Ponzi schemes.

\begin{table}
\centering
\caption{SourceP on the three metrics on the randomly divided dataset}
\begin{tabular}{llllll} 
\toprule
\textbf{Method} & \textbf{Recall} &  \textbf{Precision} &  \textbf{F-score}\\
\midrule
SourceP & 0.90 & 0.92  & 0.91 \\
\bottomrule
\end{tabular}
\label{tbl:table4}
\end{table}

\section{Related Work}
\subsection{Ponzi schemes on the blockchain}
The Ponzi scheme is a classic financial fraud \cite{38,39}. With the development of the Internet, online ``High-Yield Investment Programs" (HYIP) became a typical form of Ponzi scheme \cite{40}. Because blockchain technology became increasingly popular, unscrupulous individuals began to deploy such HYIPs on the blockchain, and more and more Ponzi schemes and Internet scammers emerged on the blockchain \cite{43}. Chen et al. \cite{42,98} first proposed machine learning-based identification of Ponzi schemes in Ethereum smart contracts by extracting features from user accounts and the opcodes of smart contracts, and then building a classification model to detect potential Ponzi schemes as smart contract implementations. Fan et al. \cite{6} based on CatBoost \cite{81}, propose the Al-SPSD model to detect newly deployed smart Ponzi schemes from the runtime opcode level promptly. Chen et al. \cite{80} propose SADPonzi, a working on Ethereum smart contract bytecode prototype system that identifies smart Ponzi schemes. Zheng et al. \cite{5} proposed building a multi-view cascade model (MulCas) to identify Ponzi schemes. Liang et al. \cite{104} detect Ponzi schemes through contract runtime behavior graph (CRBG). Ponzi schemes in blockchain platforms are not only present in Ethereum, some works have also focused on Ponzi schemes in Bitcoin and detected them \cite{41,8,44,54}.

\subsection{Smart Contract Analysis}
Many smart contract analysis studies have been conducted on smart contract security vulnerabilities in Ethereum, for example, Oyente \cite{50}, Osiris \cite{51}, Mythril \cite{52}, Maian \cite{53}, and Manticore \cite{55} use symbolic execution for vulnerability detection. Securify \cite{56} and Zeus \cite{57} use formal verification for vulnerability detection, and Slither \cite{58} and SmartCheck \cite{2} propose to use static analysis for vulnerability detection. ContractFuzzer \cite{59}, and ReGuardp \cite{60} propose vulnerability detection using fuzzy testing. Thomas et al.\cite{61} provide an empirical review of these automated analysis tools for smart contracts. Pinna et al.\cite{67} conduct a comprehensive empirical study of smart contracts deployed on the Ethereum blockchain, to provide an overview of the characteristics of smart contracts. In addition, there are empirical studies on the code smell of smart contracts \cite{62}, studies on the gas of Ethereum smart contracts \cite{63,64}, and the cloning behavior of the code \cite{65,66}.

\subsection{Pre-Trained Models for Programming Languages}
The emergence of pre-trained models (PTMs) has brought NLP into a new era \cite{68}. Some models such as BERT \cite{34} and GPT \cite{70} have recently achieved great success and have become a milestone in the field of artificial intelligence \cite{69}. Several works have also explored the application of pre-trained models (PTMs) on programming languages, such as Roberta \cite{71}, a model pre-trained on a text corpus with a Mask Language Model (MLM) learning goal, where RoBERTa(code) is pre-trained on code only. CuBERT \cite{75} is the first work to propose CodeBERT \cite{37} pre-trained on code-text pairs with MLM and replacement token detection learning goals, and is a representative pre-trained model for multilingual code representation. The major difference between GraphCodeBERT and CodeBERT is the inclusion of AST information \cite{33}. UniXcoder \cite{72} uses a masked attention matrix with prefix adapters to control the behavior of the model and enhances the code representation with cross-modal content such as AST and code annotations.PLBART \cite{77} is an application of the programming language BART \cite{78}, which incorporates the advantages of both the bidirectional encoder of the BERT model and the unidirectional left-to-right decoder in GPT. CodeT5 \cite{79} is a unified pre-trained Transformer model that makes better use of code syntax information. Large-scale pre-trained models have also evolved rapidly for code tasks, such as AlphaCode \cite{73} which uses the encoder-decoder architecture, Code-GPT \cite{76} which uses a 12-layer transformer decoder model, and GPT-C \cite{74} which is designed for the task of Code Completion.

\section{Conclusion and future work}
In this paper, we propose a method called SourceP to detect smart Ponzi schemes on Ethereum. To the best of our knowledge, this is the first detection method that uses only the source code of smart Ponzi schemes as features, and the first method to detect smart Ponzi schemes based on pre-trained models and data flows. In detecting smart Ponzi schemes on Ethereum, we experimentally demonstrate that SourceP achieves better performance and sustainability compared to existing state-of-the-art methods. We also design ablation experiments to examine the contribution of pre-trained models and data flows in SourceP. Finally, we experimentally demonstrate that SourceP possesses a good generalization capability. We explore the feasibility of using variable dependencies in source code to detect smart Ponzi schemes, avoiding some of the drawbacks of traditional smart Ponzi scheme detection methods. We reveal the potential for exploring Ponzi scheme features from smart contract source code and getting better recognition. We believe that detecting smart contracts as soon as possible after they are deployed can effectively reduce the financial losses from Ponzi schemes on Ethereum and maintain a healthy ecology of the blockchain community.
We exposing the dataset and source code we used to aid future research in this direction.

In our future work, the first step is to expand the dataset, which still has few labeled smart Ponzi schemes, and expanding the dataset has a great effect on improving the model. As mentioned in this paper, new types of smart Ponzi schemes are more difficult to detect, so more data on new types of smart Ponzi schemes are urgently needed. Or construct a Ponzi scheme dataset consisting of multiple smart contracts, to detect whether the business logic consisting of multiple contracts is a scheme. Given the excellent performance shown by pre-trained models and data flow in smart Ponzi scheme detection, we will try to explore the use of the method for other blockchain security tasks to advance blockchain security technology.

\bibliographystyle{IEEEtran}
\bibliography{IEEEabrv,base}

\end{document}